# Magnetic light

Igor I. Smolyaninov, Jill Elliott*, and Anatoly V. Zayats*

*Department of Electrical and Computer Engineering, University of Maryland, College Park, MD 20742, USA*

*\*Department of Pure and Applied Physics, Queen's University of Belfast, Belfast BT7 1NN, United Kingdom*

## Abstract

In this paper we report on the observation of novel and highly unusual magnetic state of light. It appears that in small holes light quanta behave as small magnets so that light propagation through such holes may be affected by magnetic field. When arrays of such holes are made, "magnetic light" of the individual holes forms novel and highly unusual two-dimensional "magnetic light" material. "Magnetic light" may soon become a great new tool for quantum communication and computing.

Amazing properties of metal nanowires and nanoholes are the topic of considerable current interest, due to the strong drive towards development of a new generation of nano-devices for nanotechnology, communications and computing. While studying nanoscale science of these systems, researchers constantly meet new surprising phenomena caused by yet undiscovered quantum mechanical effects. One of such recent surprises was the observation of anomalously high optical transmission of an array of nanoholes in a metal film by Ebbesen et al [1], which has been explained by light coupling to different kinds of surface plasmons on the interfaces of a metal film (surface plasmons (SPs) are the collective excitations of conductivity electrons close to the film surface and the electromagnetic field [2]). Another recent surprise was the fact that such surface-plasmon-assisted transmission does not destroy photon entanglement [3], which makes nanohole arrays very interesting for optical quantum computing applications. Since any quantum computer needs some interaction between the qubits, nonlinear interactions of surface plasmons in nanohole arrays become a topic of extreme interest for many fields of science and technology. Strong evidence of nonlinear optical processes which indeed occur inside the nanoholes due to excitation of localized SPs and cylindrical surface plasmons (CSPs) has been obtained in our recent observations of single-photon tunneling [4] and light-controlled photon tunneling effects [5]. In this article we report on our rather surprising new theoretical and experimental findings which indicate that light in nanohole arrays becomes magnetic. Although photons carry angular momentum, they can not carry magnetic moment in vacuum. This fact may not be true in every medium. For example, recently demonstrated symmetry between electrons and the electromagnetic modes of metal nanowires and nanoholes (cylindrical surface plasmons) strongly suggests that plasmons with nonzero angular momenta may have nonzero magnetic moments as well. Here we show that such cylindrical plasmons can indeed exhibit magnetization on the nanoscale, and thus, be sensitive to the external magnetic field. Such an unusual "magnetic" state of light which is of great value in quantum communication applications has been observed in our experiments on the magnetic-field-induced changes in the radiation spectrum of cylindrical surface plasmons excited in an array of cylindrical nanoholes. It seems that the nonlinear optical coupling between the cylindrical plasmons of different nanoholes in the array may be described as magnetic interactions in a two-dimensional lattice of magnetic moments reminiscent of the Ising model. Magnetization of light in such a nanohole array has been observed directly using magnetic force microscopy.

Cylindrical surface plasmons which can exist on a cylindrical surface of the metallic wire (or hole in a metal film) are described by the wave vector $k_z$ related to CSP propagation in the axial direction of the infinite cylinder and the angular quantum number $n$ related to the azimuthal CSP propagation ($k_\psi$) along the cylinder circumference (Fig. 1a). As a result, the CSP trajectory on a cylindrical surface can be imagined as a spiral with the period determined by CSP angular quantum number: For $n=0$ there is no angular momentum and such CSP is analogous to the surface plasmon on a plane surface propagating on the cylinder wall in the axial direction, while for $n>>1$ CSPs are strongly rotating around the cylinder and their spectrum converges quazi-continuously to $\omega_p/2^{1/2}$ (where $\omega_p$ is the plasma frequency of the metal) in the large wave vector limit. While linear optics of cylindrical surface plasmons is relatively well understood [6], until recently there was no theoretical description of their nonlinear optics. Surprisingly, it appears that the theory of CSP interaction [7] has a lot in common with the description of electrodynamics of electric charges in the class of high energy physics theories called the Kaluza-Klein theories, which introduce compact and small extra spatial dimensions.

Similarity between the CSPs and the electric charges stems from the way in which the electric charges are introduced in the original five-dimensional Kaluza-Klein theory (see for example [8]). In this theory the electric charges are introduced as chiral (nonzero angular momentum $n \neq 0$) modes of a massless quantum field, which is quantized over the cyclic compactified fifth dimension. The electric charge of each field mode is proportional to the fifth component of this mode's momentum $n$. Thus, the electric charge conservation becomes a simple consequence of the momentum conservation law. In an analogous way, cylindrical surface plasmons of a nanowire or a nanohole may be considered as if they exist in a curved three-dimensional space-time defined by the metal interface, which besides an extended $z$-coordinate, has a small "compactified" angular $\psi$–dimension along the circumference of the cylinder (Fig.1b). As a result, the theory of CSP mode propagation and interaction may be formulated as a three-dimensional Kaluza-Klein theory, and similarity between the effective CSP chiral charge (proportional to the angular momentum $n$ of the CSP) and the normal electric charge becomes evident: both kinds of charges are proportional to the respective quantized component of the angular momentum $n$. As a result, we arrive to a physical picture of CSP interaction [7] in which the higher ($n>0$) CSP modes posses quantized effective chiral charges proportional to their angular momentum $n$. In a metal nanowire these slow moving effective charges exhibit long-range interaction via exchange of fast massless and chargeless CSPs with zero angular momentum.

If we recall that in addition to the electric charge, electron posses a magnetic moment, we may ask ourselves if the described symmetry between electrons and the cylindrical surface plasmons implies that CSPs with nonzero angular momenta may have nonzero magnetic moments as well. This would be quite an unusual and interesting situation since normally photons in vacuum do not carry the magnetic moment. If the answer is yes, we would deal with a new form of "magnetic" light state with quite exotic properties: individual quanta of the CSP electromagnetic field would behave like small magnets. The magnetic moment $\mu$ of the individual CSP quanta can be estimated as

$$\mu = -\hbar\left(\frac{\partial \omega}{\partial H}\right)_{H=0} = -\frac{ea^2}{c}\left(\frac{\partial \omega}{\partial \phi}\right)_{\phi=0} \qquad (1)$$

where $H$ is the external magnetic field applied in the axial direction, $a$ is the cylinder radius, and $\phi$ is the number of magnetic flux quanta inside the cylinder. This estimation is easy to perform in the two limiting cases $n=0$ and $n\rightarrow\infty$ described above if the nonretarded electrostatic approximation is used [9] in which surface plasmons have zero dispersion $d\omega/dk$. These cases correspond to the surface plasmons with large wave vectors, which propagate parallel and perpendicular to the axial magnetic field, respectively. If we neglect the curvature of the cylinder, the results of [9] are reproduced: for $n=0$ CSPs $\omega = (\omega_p{}^2 + \omega_c{}^2)^{1/2}/2^{1/2}$, and for $n\neq 0$ CSPs $\omega = (\omega_p{}^2 + \omega_c{}^2)^{1/2}/2^{1/2} \pm \omega_c/2$, where $\omega_c=eH/mc$ is the cyclotron frequency, and the sign in front of the $\omega_c/2$ term is determined by the rotation direction. Thus, it is evident that the $n=0$ plasmons do not have magnetic moments, while $n\neq 0$ CSP quanta have magnetic moments of the order of Bohr's magneton $\mu_B = e\hbar/2mc$. Exact values of the CSP magnetic moments can be determined by taking into account the effects of curvature of the cylinder, retardation, and the Aharonov-Bohm effect [10,11] on the CSP dispersion curve. Analysis of the numerical calculations of the CSP dispersion presented in [10,11] for a number of different geometries indicates that the results of our simple estimate remains valid in all these cases: CSPs with zero angular momentum have no magnetic moment, while $n \neq 0$ CSPs have $\mu \sim \mu_B$. As a result, although rather weak, magnetism of cylindrical surface plasmons should be detectable in the experiment.

Magnetic properties of CSPs should manifest themselves in the optical properties of cylindrical plasmons in the external magnetic field. Influence of the applied dc magnetic field on cylindrical surface plasmons possessing magnetic moments will lead to the change of the radiation spectrum associated with CSPs. Since cylindrical surface plasmons with large $n>>1$ have dispersion curves approaching, in the limit of large wave vectors, the surface plasmon frequency, it is these CSPs that will contribute most to the magnetic field effects near $\omega = \omega_p / 2^{1/2}$ due to their large magnetic moment and high density of states (since $d\omega/dk$ is small for $k \rightarrow \infty$). However, such CSPs cannot be excited directly by the incident light because of the difference in their wave vectors. A periodic array of cylindrical holes facilitates the excitation of CSPs in this spectral range due to diffraction effects.

The measurements of the effect of magnetic field on the spectrum of cylindrical surface plasmons (Fig.2) strongly indicate the suggested magnetic behavior. The transmission spectra of a periodic array of cylindrical holes (inset in Fig.2) created in a 40 nm thick gold film on a glass substrate using focused ion beam milling have been measured with and without applied 0.46 T magnetic field having component perpendicular to the film (parallel to the holes axis). The transmission spectrum of the array measured under the white light illumination without the applied magnetic field has two major features in the spectral range of plasmon excitation. The first one is the peak of anomalous light transmission around 500 nm light wavelength, which corresponds to the excitation of surface plasmons on the gold-air interface. This feature is similar to the peak of anomalous light transmission observed in [1].  The second spectral feature close to 400 nm (which is just above the surface plasmon frequency) should at least partially correspond to the excitation and re-radiation of cylindrical surface plasmons with large angular momenta in the holes. In this spectral range, the double period structure provides the excitation of a set of surface plasmon polariton modes on a gold-air interface corresponding to higher in-plane diffraction orders which can then couple to CSPs. This second spectral feature is observed to depend on the applied magnetic field, while the first one shows no sensitivity. This weak magnetic field dependence corresponds well to the expected weak dependence of CSP radiation on the applied magnetic field.

It is also interesting to note that the nonlinear optical coupling between the plasmons of different nanoholes in the array (which may be important in such applications as quantum communication and computing) may be described as magnetic interactions in a two-dimensional lattice of magnetic moments. Although CSPs are bosons and do not exhibit strong exchange interaction, due to their statistics they would "like" to be in the same magnetic state when excited by circular polarized coherent light. As a result, a two-dimensional lattice of magnetic moments (inset in Fig.2(a)) reminiscent of the Ising model will be formed. Thus, novel two-dimensional "photonic" magnetic materials may be introduced.  Such "magnetic light" state in a nanohole array should be possible to observe directly using magnetic force microscopy (MFM). However, local detection of relatively weak magnetism of light in the nanohole array (our estimates indicate magnetic moments of no larger than $10^4$ $\mu_B$ under illumination with 100 mW of the CW 488 nm laser light) requires sacrifice of the MFM spatial resolution for the sake of sensitivity. Thus, a custom MFM has been built (Fig.3c), which uses larger than usual (made of 50 μm radius sharpened Ni wire), and mechanically softer (~$10^{-3}$ N/m force constant) magnetic tips. Images of the local magnetization of a 30x30 μm$^2$ array of nanoholes coated with thin polydiacetilene film (similar to the one described in detail in [5]) obtained under illumination with 488 nm circular polarized light are presented in Fig.3(a,b). Polymer coating shifts both cylindrical surface plasmon (CSP) and the regular surface plasmon (SP) resonances up in the wavelength, due to the large dielectric constant of the polymer coating [2]. Thus, the CSP resonance shifts conveniently with respect to the spectrum in Fig.2a to the wavelength range of the set of Argon ion laser lines used in the experiment. Characteristic bright and dark stripes indicating magnetic dipole behavior of the entire illuminated nanohole array and some smaller structures within the array are clearly visible in the lower portions of the images in Fig.3(a,b). The observed magnetization signal exhibit rather striking sharp spectral behavior. The image in Fig.3(d) measured outside the CSP resonance under the illumination with 514 nm light over the same area as in Fig.3(a) shows no magnetization signal. Thus, direct and clear evidence of the magnetic behavior of light in the nanohole array has been obtained.

In addition to the importance of our results to fundamental nanoscience, our observations suggest the possibility to control transmission of nanoholes at a single-photon level with an external magnetic field. This possibility is extremely attractive in quantum communication applications.

This work has been supported in part by the NSF grant ECS-0210438

**Figure Captions**

Fig.1 (a) Geometry of the cylindrical surface plasmon on a surface of a metal nanowire or nanohole in a metal film. (b) Symmetry between a cylindrical surface plasmon and an electron: in Kaluza-Klein theories electron is represented as a chiral mode of a massless field which is quantized over an additional compactified spatial coordinate.

Fig.2 (a) Optical transmission T(0) of the nanohole array without applied magnetic field. (b) Spectrum of the relative change (T(H)-T(0))/T(0) of the optical transmission in the applied magnetic field. The insert in (a) shows the electron microscope image of the double period nanostructure: the hole diameter is 150 nm, periodicity is $D_{1,2}$ = 600 nm and 1200 nm. The magnetic moments of cylindrical plasmons are shown by the arrows. The insert in (b) illustrates schematics of the experimental set-up: (WL) white light source, (L) lens, (P) polarizer, (S) sample, (H) applied magnetic field.

Fig.3 (a) 60x60 μm image of the light induced magnetization of a 30x30 $μm^2$ array of nanoholes (located in the bottom left corner) obtained under illumination with 488 nm circular polarized light. The 30x30 μm zoom area shown in (b) is indicated by a square boundary. (c) Schematic view of the illumination geometry. (d) Image of the same area obtained with 514 nm light shows no trace of magnetization.

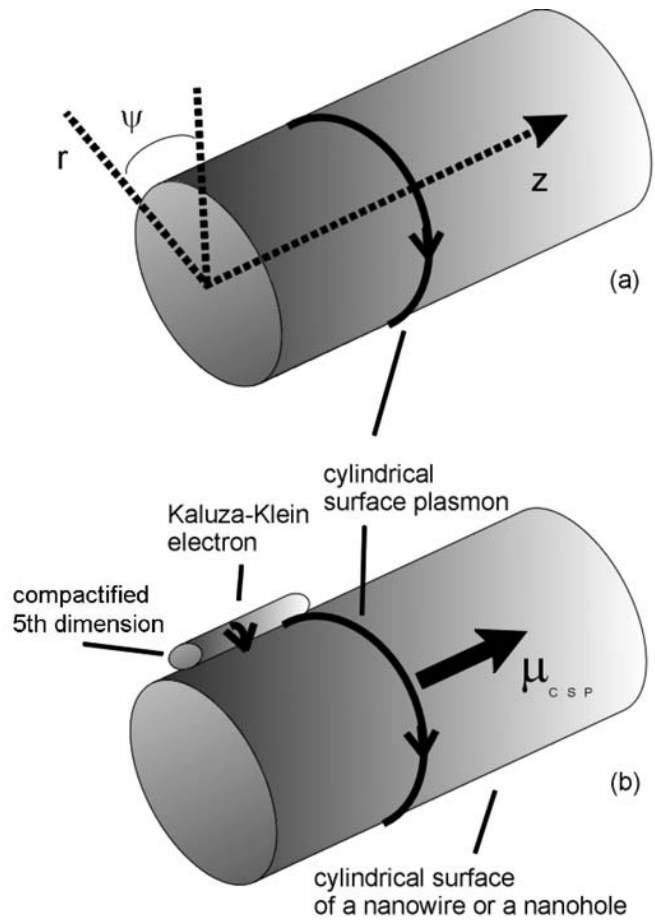

Fig.1

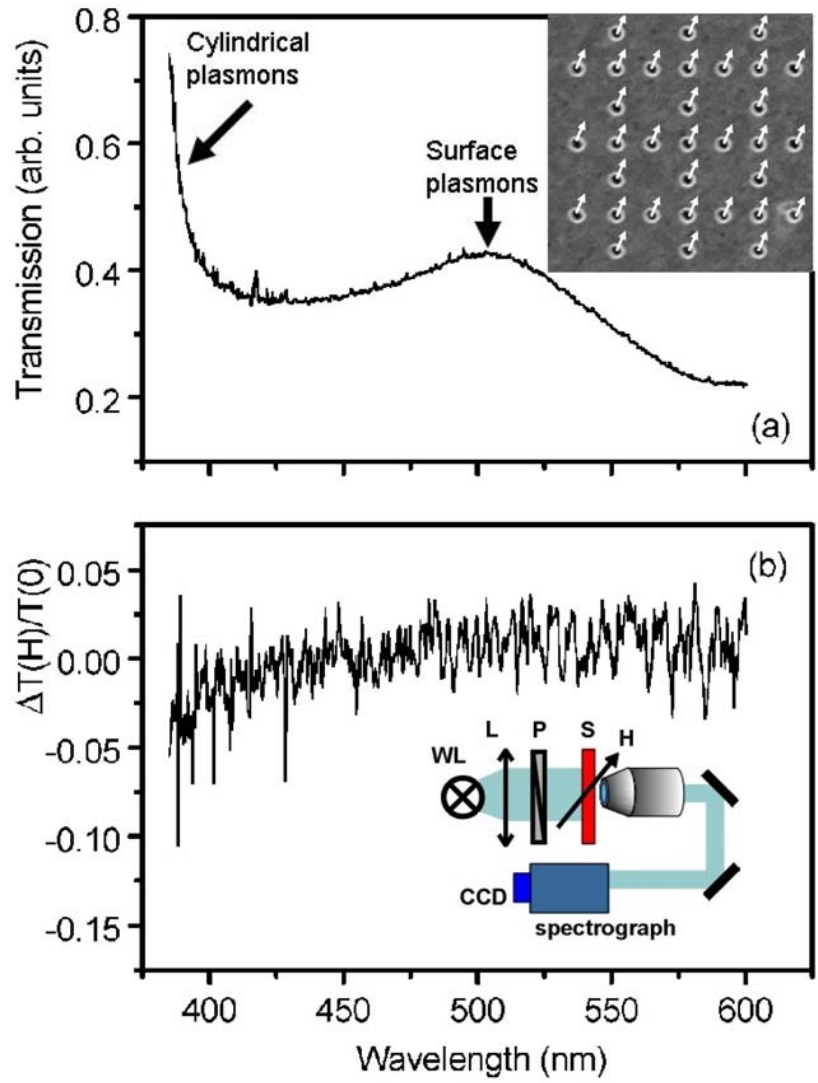

Fig.2

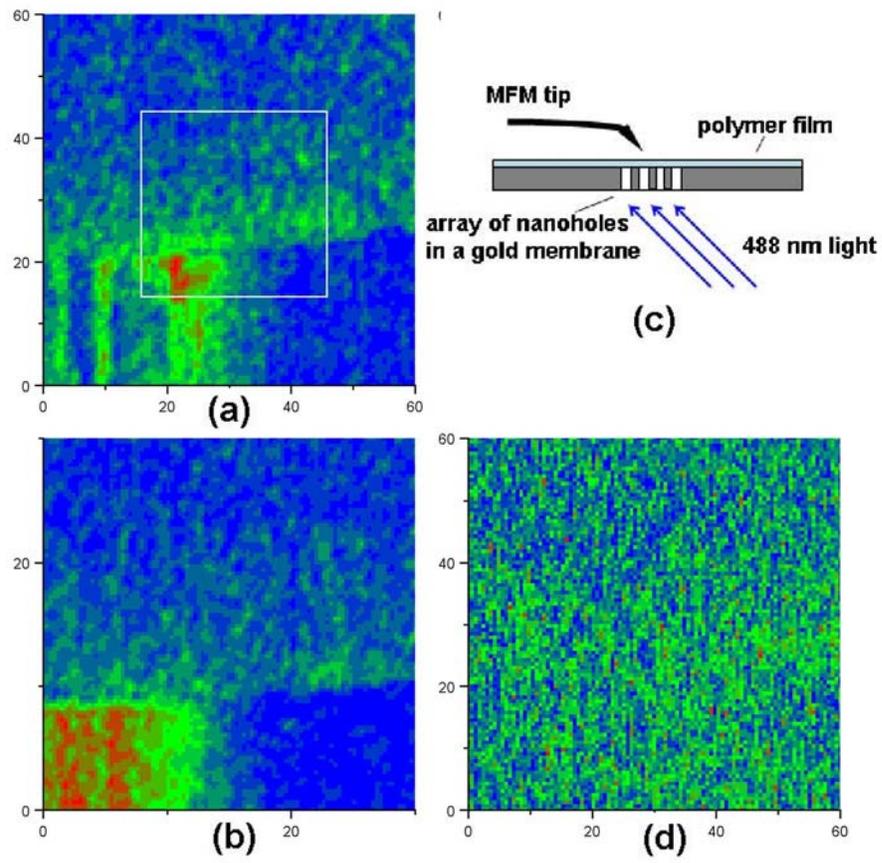

Fig.3